\documentclass[aps,prl,twocolumn,superscriptaddress,preprintnumbers,%
               showpacs,nofootinbib,floatfix]{revtex4}
\newcommand{\PRE}[1]{}       % Use if journal style
\usepackage{bm}
\usepackage{epsfig}
\usepackage{hyperref}

\newcommand{\postscript}[2]{\setlength{\epsfxsize}{#2\hsize}
   \centerline{\epsfbox{#1}}}

\def\eslt{\not\!\!\!{E_T}}
\def\to{\rightarrow}

\def\bi{\begin{itemize}}
\def\ei{\end{itemize}}

\def\tb{\tilde b}

\def\tst{\tilde t}

\def\tg{\tilde g}

\def\tq{\tilde q}
\def\tw{\widetilde W}
\def\tz{\widetilde Z}

\def\mtmin{m_T^{\rm min}}
\def\alt{\lesssim}
\def\agt{\gtrsim}
\def\be{\begin{equation}}  
\def\ee{\end{equation}}  
\def\bea{\begin{eqnarray}}  
\def\eea{\end{eqnarray}}

\newcommand\plb[3]{{\it Phys\ Lett.\ }{\bf B #1} (#2) #3}
\newcommand\jhep[3]{{\it J.\ High Energy Phys.\ }{\bf #1} (#2) #3}
\newcommand\npb[3]{{\it Nucl.\ Phys.\ }{\bf B #1} (#2) #3}

\newcommand\prD[3]{{\it Phys.\ Rev.\ }{\bf D #1} (#2) #3}
\newcommand\prL[3]{{\it Phys.\ Rev. \  Lett. \ }{\bf #1} (#2) #3}

\begin{document}

\preprint{OU-HEP-130215, UH-511-1207-13}

\title{
\PRE{\vspace*{1.5in}}
Same sign diboson signature from supersymmetry models \\
with light higgsinos at the LHC
%from radiative natural supersymmetry at the LHC
\PRE{\vspace*{0.3in}}
}
\author{Howard Baer}
\affiliation{Dept. of Physics and Astronomy,
University of Oklahoma, Norman, OK, 73019, USA
\PRE{\vspace*{.1in}}
}
\author{Vernon Barger}
\affiliation{Dept. of Physics,
University of Wisconsin, Madison, WI 53706, USA
\PRE{\vspace*{.1in}}
}
\author{Peisi Huang}
\affiliation{Dept. of Physics,
University of Wisconsin, Madison, WI 53706, USA
\PRE{\vspace*{.1in}}
}
\author{Dan Mickelson}
\affiliation{Dept. of Physics and Astronomy,
University of Oklahoma, Norman, OK, 73019, USA
\PRE{\vspace*{.1in}}
}
\author{Azar Mustafayev}
\affiliation{Dept. of Physics and Astronomy,
University of Hawaii, Honolulu, HI 96822, USA
\PRE{\vspace*{.1in}}
}
\author{Warintorn Sreethawong}
\affiliation{School of Physics,
Suranaree University of Technology, 
Nakhon Ratchasima 30000, Thailand
\PRE{\vspace*{.1in}}
}
\author{Xerxes Tata}
\affiliation{Dept. of Physics and Astronomy,
University of Hawaii, Honolulu, HI 96822, USA
\PRE{\vspace*{.1in}}
}

%\date{October 15, 2011}

\begin{abstract}
\PRE{\vspace*{.1in}}

In supersymmetric models with light higgsinos (which are motivated by
electroweak naturalness arguments), the direct production of higgsino pairs may be 
difficult to search for at LHC due to the low visible energy release
from their decays.  
However, the wino pair production reaction $\tw_2^\pm\tz_4\to (W^\pm\tz_{1,2})+(W^\pm\tw_1^\mp)$ 
also occurs at substantial rates and leads to final states including 
equally opposite-sign (OS) and same-sign (SS) diboson production.
We propose a novel search channel for LHC14 based on the SS diboson
plus missing $E_T$  final state which contains only modest jet activity. 
Assuming gaugino mass unification, and an integrated luminosity $\agt 100$~fb$^{-1}$, 
this search channel provides a reach for SUSY well beyond that from usual gluino pair production.

\end{abstract}

%12.60.-i   Models beyond the standard model
%95.35.+d   Dark matter

\maketitle

The recent discovery of a Higgs-like resonance at $m_h\sim 125$~GeV by
the Atlas and CMS collaborations\cite{atlas_h,cms_h} completes the
identification of all the states in the
Standard Model (SM). 
However, the existence of fundamental scalars in the SM is problematic in that
they lead to gauge instability and fine-tuning issues. 
Supersymmetric (SUSY) theories stabilize the scalar sector due to a
fermion-boson symmetry, thus providing a solution to the gauge hierarchy problem\cite{witten}. 
In fact, the measured Higgs boson mass
$m_h\simeq 125$~GeV falls squarely within the narrow range
predicted\cite{mhiggs} by the minimal supersymmetric Standard Model
(MSSM); this may be interpreted as indirect support for weak scale SUSY.
In contrast, the associated superparticle states have failed to be
identified at LHC, leading the Atlas and CMS
collaborations\cite{atlas_s,cms_s} to place limits of $m_{\tg}\agt 1.4$~TeV 
(for $m_{\tg}\simeq m_{\tq}$) and $m_{\tg}\agt 0.9$~TeV (for
$m_{\tg}\ll m_{\tq}$) within the popular mSUGRA/CMSSM
model\cite{msugra}. 
%More generally, it is the first generation
%squark masses that enter most directly in the LHC analyses.

\begin{figure}[tbp]
\postscript{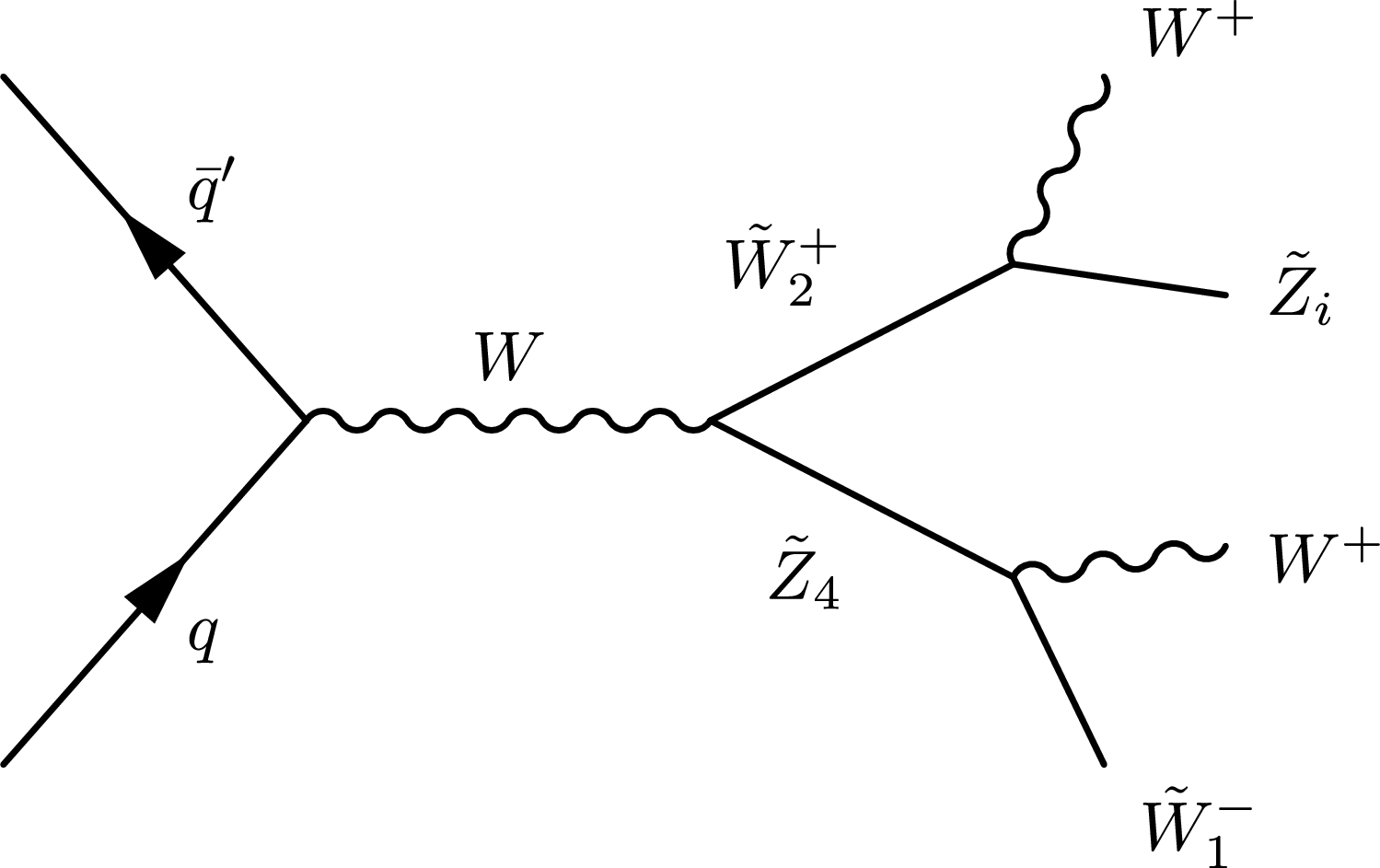}{0.9}
\caption{Diagram depicting same-sign diboson production at LHC in SUSY models 
with light higgsinos.
\label{fig:diagram}}
\end{figure}

In many SUSY models used for phenomenological analyses, the
higgsino mass parameter $|\mu|$ is larger than the gaugino mass
parameters $|M_{1,2}|$. In the alternative case where $|\mu| \ll
|M_{1,2}|$, the lighter electroweak chargino $\tw_1$ and the lighter
neutralinos $\tz_{1,2}$ are higgsino-like, while (assuming $|M_2|>|M_1|$)
the heavier chargino and the heaviest neutralino $\tz_4$ is wino-like,
and $\tz_3$ is bino-like. Electroweak $\tw_2\tz_4$ 
production which occurs with $SU(2)$ gauge strength then leads to a
novel $W^\pm W^\pm + \eslt$ signature via the process
shown in Fig.~\ref{fig:diagram}.
We examine prospects for observing this signal in the 14~TeV run
of the CERN LHC.

Models with light higgsinos have a number of theoretical advantages, and
have recently received considerable attention. To understand why, we
note that the
minimization condition for the Higgs scalar potential leads
to the well known (tree-level) relation, 
\be 
\frac{M_Z^2}{2} =
\frac{m_{H_d}^2-m_{H_u}^2\tan^2\beta}{(\tan^2\beta -1)} -\mu^2
\simeq -m_{H_u}^2-\mu^2
\label{eq:mssmmu}
\ee 
where $m_{H_u}^2$ and $m_{H_d}^2$ are the tree-level mass squared
parameters of the two Higgs doublets that are required to give masses to
up- and down-type quarks, and $\tan\beta$ is the ratio of their vacuum
expectation values. The value of $M_Z$ that is obtained from
(\ref{eq:mssmmu}) is {\em natural} if the three terms on the
right-hand-side (RHS) each have a magnitude of the same order as
$M_Z^2$, implying $\mu^2/(M_Z^2/2)$ is limited from above by the extent of
fine-tuning one is willing to tolerate. The lack of a chargino signal at
the LEP2 collider requires $|\mu|\agt 103.5$~GeV \cite{lep2ino}, so that
light higgsino models with low fine-tuning favour $|\mu|\sim
100-300$~GeV (in fact, $\mu^2$ was suggested as a measure of fine-tuning
in Ref.~\cite{ccn}). When radiative corrections to (\ref{eq:mssmmu}) are
included, masses of other superpartners (most notably third generation
squarks) also enter on the RHS, and large cancellations may be needed if
these have super-TeV masses.
% exacerbating what has been referred to as
%the {\em little hierarchy problem}.
% how do large values of SUSY model
%parameters conspire to give a $Z$-boson or Higgs boson mass around just
%100~GeV?  This led to the examination of models that favour values of
%$|\mu|$ that are not hierarchically larger\footnote{Small $|\mu|$ is a
%necessary, but not sufficient condition for there to be no large
%cancellations in (\ref{eq:mssmmu}).} than $M_Z$ 
Models favouring low values of $|\mu|$ include:
%
%While SUSY solves the big hierarchy problem, the required super-TeV
%sparticle masses seemingly exacerbate the so-called ``little hierarchy
%problem'': how do such large values of SUSY model parameters conspire to
%yield a $Z$ boson (or Higgs boson) mass of just 91.2 (125)~GeV? Such
%considerations have led to renewed scrutinization of what it means for
%SUSY theories to obey electroweak naturalness.  Minimization of the
%tree-level MSSM scalar potential in the neutral Higgs field directions
%yields the well-known relation that
%
%\be
%\frac{m_Z^2}{2} = \frac{m_{H_d}^2-m_{H_u}^2\tan^2\beta}{(\tan^2\beta -1)}
%-\mu^2\simeq -m_{H_u}^2-\mu^2 
%\label{eq:mssmmu},
%\ee where the latter approximate equality obtains for ratio-of-Higgs
%vevs $\tan\beta\equiv v_u/v_d\agt 3$.  To be quantitative, a finetuning
%measure $\Delta_{EW} =max_i\left[C_i/(m_Z^2/2)\right]$ may be defined,
%where $C_i$ represents any of the terms on the right-hand-side of
%Eq. \ref{eq:mssmmu}.  Models with $\Delta_{EW} <100$ would then
%correspond to better than $\Delta_{EW}^{-1}=1\%$ electroweak finetuning
%(EWFT).  To enjoy low values of EWFT, both $\mu^2 $ and $m_{H_u}^2$ in
%Eq. \ref{eq:mssmmu} should be of order $m_Z^2/2$.  In fact, in
%Ref. \cite{ccn}, the $\mu$ parameter itself was recommended as a measure
%of EWFT.  Several recent SUSY models which include low $|\mu |\sim m_Z$
%include: 
%
\bi
\item the hyperbolic branch/focus point (HB/FP) region of minimal
supergravity model (mSUGRA or CMSSM)\cite{hb_fp} or its non-universal Higgs mass
extension\cite{fs},
\item models of ``natural SUSY'' (NS)\cite{kn,ah,ns,nat} which have
$\mu\sim 100-300$~GeV, top- and bottom-squarks with
$m_{\tst_{1,2}},\ m_{\tb_1}\alt 500$~GeV and $m_{\tg}\alt 1.5$~TeV, and
\item radiative natural SUSY (RNS)\cite{rns}, where again $\mu\sim
100-300$~GeV and where $m_{H_u}^2$ is driven to small values $\sim
-M_Z^2$ via the large top quark Yukawa coupling.  
\ei 
The HB/FP region of mSUGRA\cite{msugra} remains viable\cite{dm125} but suffers high fine-tuning due to large top squark masses. 
The NS models as realized within the
MSSM also seem to be disfavoured because much heavier top-squark masses are
required to lift $m_h$ up to 125~GeV and to bring the $b\to s\gamma$
branching fraction into accord with measurements\cite{nat}. Models of NS
with extra exotic matter which provide additional contributions to $m_h$
would still be allowed\cite{hpr}.  The RNS model allows for top- and
bottom-squarks in the 1-4~TeV range, and with large mixing can
accommodate $m_h\simeq 125$~GeV and $BF(b\to s\gamma )$ while
maintaining cancellations in (\ref{eq:mssmmu}) at the 3-10\% level.

Another potential advantage of models with light higgsinos is that if 
the lightest supersymmetric particle (LSP) is higgsino-like, then it
annihilates rapidly in the early universe, thus avoiding cosmological
overclosure bounds. In this case, the higgsino might serve as a co-dark-matter
particle along with perhaps the axion\cite{az1}.

Although the production of charged and neutral higgsinos may occur at large
rates (pb-level cross sections for $\mu \sim 150$~GeV at the LHC), detection of these
reactions is very difficult because the
mass gaps $m_{\tw_1}-m_{\tz_1}$ and $m_{\tz_2}-m_{\tz_1}$ are typically small, $\sim 5-20$~GeV, resulting in very low visible energy release from
$\tw_1$ and $\tz_2$ decays. Thus, higgsino pair production events are
expected to be buried beneath SM backgrounds\cite{bbh}. 
We examine instead signals from the heavier
gaugino-like states focusing on the wino-like states $\tw_2$ and $\tz_4$,
whose production
cross sections will be fixed by essentially just the wino mass parameter
$M_2$ if first generation squarks are heavy.
% and $M_2$ is large enough
%so that $\tw_2$ and $\tz_4$ are mainly winos.

%If $|\mu |$ is much lighter than the bino mass $M_1$ or the wino mass
%$M_2$, then one expects the heavier chargino $\tw_2$ to be highly
%wino-like while $\tz_3$ and $\tz_4$ are either bino-like or
%wino-like. For simplicity of discussion, we will assume gaugino mass
%unification $M_1=M_2=M_3\equiv m_{1/2}$ at the grand unified theory
%(GUT) scale $Q\simeq 2\times 10^{16}$~GeV, so that at the weak scale the
%gaugino masses are in the approximate ratio $M_1:M_2:M_3\sim 1:2:7$ (our
%forthcoming results are more general than this).  Then the LHC gluino
%search result implies $m_{\tg}\simeq M_3\agt 1.1$~TeV so that $M_2\agt
%300$~GeV and $M_1\agt 150$~GeV. Then for heavy squarks, gluino pair
%production is also suppressed at LHC, and prospects for SUSY discovery
%of such models at LHC might appear dim. However, wino pair production
%occurs through the rather large $SU(2)$ gauge coupling, and since
%$M_2\ll M_3$, it doesn't suffer the mass suppressed cross section as in
%gluino pair production.  Thus, in the case of $\tw_2^\pm\tz_4$
%production with $\tw_2$ and $\tz_4$ both wino-like, we might expect this
%to comprise one of the dominant {\it observable} SUSY cross sections.

%To provide concrete results, 
As an illustration, 
we show sparticle production cross sections
for a model line from the RNS model, which can be generated
from the two-extra-parameter non-universal Higgs model
(NUHM2)~\cite{nuhm2} with parameters 
\be 
m_0,\ m_{1/2},\ A_0,\ \tan\beta,\ \mu\ \ \ {\rm and}\ \ m_A .  
\ee
The independent GUT
scale parameters $m_{H_u}^2$ and $m_{H_d}^2$ have been traded for
convenience for the weak scale parameters $\mu$ and $m_A$.  We take
$m_0=5$~TeV, $A_0=-1.6m_0$, $\tan\beta =15$, $\mu =150$~GeV, $m_A=1$
TeV, and allow $m_{1/2}$ to vary between $300-1000$~GeV. The
large negative $A_0$ value allows $m_h\sim 125$~GeV\cite{h125} and at
the same time limits the cancellation between the terms in
(\ref{eq:mssmmu}) to no better than 3.5\%.
%the EWFT $\Delta_{EW}$ varies from $16$ to $28$ as $m_{1/2}$ increases
%(6\% to 3.5\% EWFT).  
%
We use Isajet\cite{isajet} for spectrum generation, branching fractions
and also later for signal event generation.

\begin{figure}[tbp]
\postscript{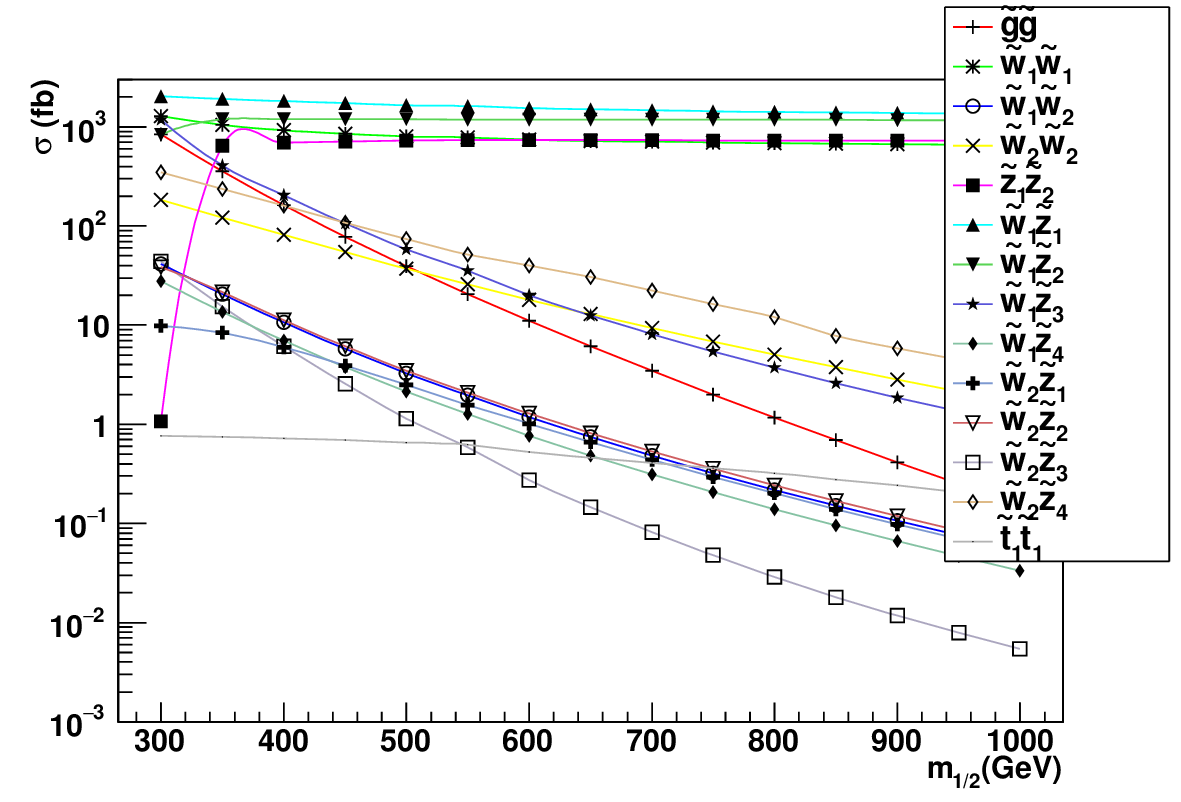}{1.0}
\caption{Plot of various sparticle pair production cross sections from
the RNS model line at LHC with $\sqrt{s}=14$~TeV.
\label{fig:xsec}}
\end{figure}

The cross sections for various electroweak-ino pair production are shown
versus $m_{1/2}$ in Fig.~\ref{fig:xsec} for $pp$ collisions at
$\sqrt{s}=14$~TeV, where we have used Prospino\cite{prospino} to
obtain results at next-to-leading-order in QCD.  The difficult-to-detect
$\tw_1^+\tw_1^-$, $\tw_1\tz_1$ and
$\tz_1\tz_2$ higgsino processes dominate  sparticle production with a
cross section
$\sigma\sim 1000$~fb. The corresponding
curves are nearly flat with $m_{1/2}$
variation since $\mu$ is fixed at $150$~GeV.
%
%These reactions as
%mentioned are difficult to see at LHC since the visible decay products
%are rather soft. 
%
The charged and neutral wino-like states $\tw_2$ and $\tz_4$ are mainly
produced via $\tw_2\tz_4$ and $\tw_2^+\tw_2^-$ reactions with cross
sections that begin at $\sim 400$~fb but fall slowly with increasing
$m_{1/2}$ because their masses increase with $m_{1/2}$ (since
$m_{\tw_2}\simeq m_{\tz_4} \simeq M_2\sim 0.8 m_{1/2}$). 
%$SU(2)$ gauge
%symmetry precludes a coupling of neutral wino pairs to $Z$; as a result
%the $\tz_4\tz_4$ production rate is tiny in the limit of heavy (first
%generation) squarks.  
Cross sections for mixed gaugino-higgsino
production reactions such as $\tw_2\tz_2$, $\tw_1\tz_3$ {\it etc.}  fall
more rapidly with $m_{1/2}$ and become subdominant.  The gluino pair
production cross section (`+'s on the red curve) starts at $\sim 1000$~fb, 
but drops rapidly as $m_{1/2}$ (alternatively $m_{\tg}\simeq 2.4m_{1/2}$) increases. 

\begin{figure}[tbp]
\postscript{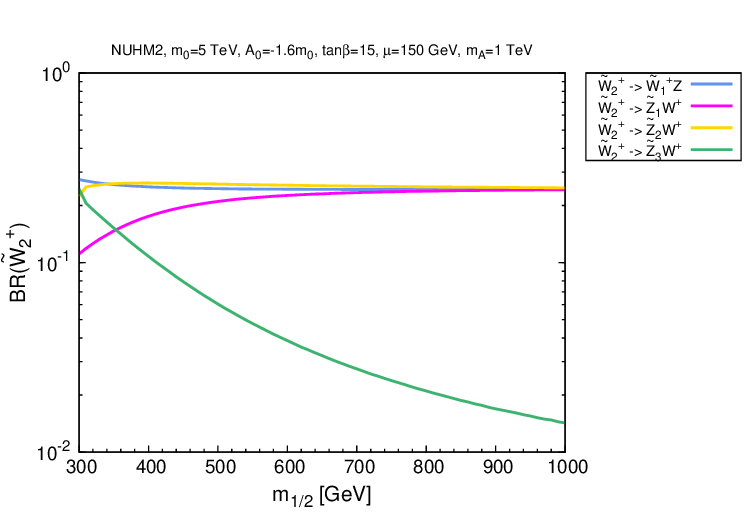}{0.9}
\caption{Branching fractions of $\tw_2$ along the RNS model line.
\label{fig:bfw2}}
\end{figure}

To understand the final states, we show in Fig. \ref{fig:bfw2} the
dominant $\tw_2$ branching fractions versus $m_{1/2}$ along the same
model line. Here, we see that $\tw_2^+\to\tw_1^+ Z$ and $\tz_2\ W^+$ at
about 25\% each while $\tw_2^+\to\tz_1 W^+$ is increasing with $m_{1/2}$
to also approach $\sim 25\%$.

\begin{figure}[tbp]
\postscript{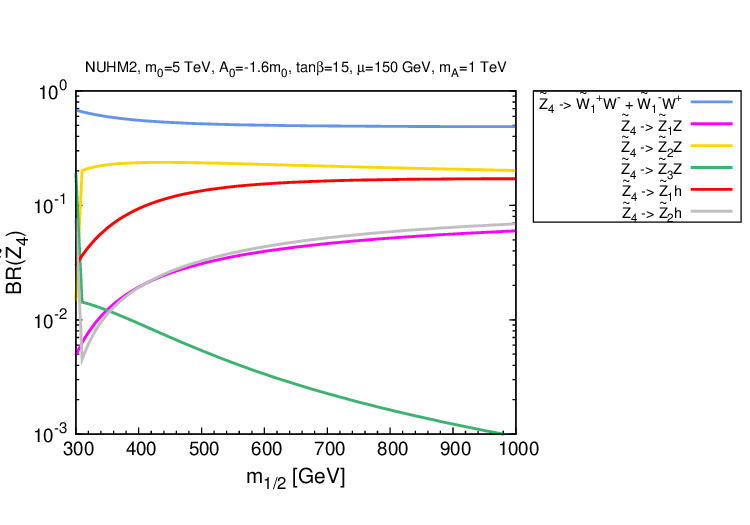}{0.9}
\caption{Branching fractions of $\tz_4$ along the RNS model line.
\label{fig:bfz4}}
\end{figure}

In Fig. \ref{fig:bfz4}, we show the $\tz_4$ branching fraction versus
$m_{1/2}$, and here find $\tz_4\to \tw_1^+ W^- +\tw_1^- W^+$ occurring
at $\sim 50\%$, followed by $\tz_4\to \tz_2 Z$ and $\tz_1 h$ occurring
at $\sim 15-20\%$ level; several other subdominant decay modes are also
shown.

Combining the $\tw_2^\pm \tz_4$ production reaction with decay modes,
the following potentially interesting signatures emerge: 
\begin{itemize}
\item $\tw_2^\pm\tz_4 \to \left(W^+W^-,\ WZ,\ ZZ\ {\rm and}\ W^\pm
W^\pm\right) +\eslt$.  
\end{itemize}
(The $W^+W^-$, $WZ$ and $ZZ$ plus $\eslt$ signals also arise from 
chargino and neutralino production in models such as mSUGRA/CMSSM.)
The $W^+W^-$ signal will likely be buried beneath prodigious SM 
backgrounds from 
$W^+W^-$ and $t\bar{t}$ production, while the $ZZ$ signal is 
likely to be rate-limited at least in the golden four lepton mode. 
There may also exist some limited LHC14
reach for the $WZ\to 3\ell$ signal as in Ref.~\cite{wz}.  
However, same-sign diboson production-- $W^\pm W^\pm +\eslt$-- 
{\it is a novel signature, characteristic of the light higgsino scenario.}  
Assuming leptonic decays of the $W$
bosons, we expect events with same-sign (SS) dileptons $+\eslt$
accompanied by modest levels of hadronic activity arising from initial
state QCD radiation and from hadronic decays of $\tw_1$ or $\tz_2$ where
the usually soft decay products might become  boosted to create
a jet.  The SS dilepton signal emerging from wino-pair
production is quite distinct from that expected from gluino pair
production\cite{ssdil} since in the latter case several very high $p_T$
jets and large $\eslt$ are also expected.  

The SM physics backgrounds to the SS diboson signal come from $uu\to
W^+W^+ dd$ or $dd\to W^-W^- uu$ production, with a cross section
$\sim 350$~fb. These events will be characterized by high rapidity
(forward) jets and rather low $\eslt$. $W^\pm W^\pm$ pairs may also occur 
via two over-lapping events; such events will mainly have low $p_T$ $W$s
and possibly distinct production
vertices. Double parton scattering will also lead to SS diboson events, at a
rate somewhat lower than the $qq \to W^\pm W^\pm q'q'$
process\cite{stirling}. Additional physics backgrounds come from
$t\bar{t}$ production where a lepton from a daughter $b$ is
non-isolated, from $t\bar{t}W$ production, and $4t$ production.  SM
processes such as $WZ\to 3\ell$ and $t\bar{t}Z\to 3\ell$ production,
where one lepton is missed, constitute {\em reducible} backgrounds to the
signal.
%, from $t\bar{t}$ production where one
%isolated lepton comes from a $b$-quark semileptonic decay or from higher
%order processes like $t\bar{t}W$ production.

To estimate background, we employ a toy detector simulation with
calorimeter cell size $\Delta\eta\times\Delta\phi=0.05\times 0.05$ and
$-5<\eta<5$ . The HCAL (hadronic calorimetry) energy resolution is taken
to be $80\%/\sqrt{E}\oplus 3\%$ for $|\eta|<2.6$ and FCAL (forward
calorimetry) is $100\%/\sqrt{E} \oplus 5\%$ for $|\eta|>2.6$, where the two
terms are combined in quadrature. The ECAL (electromagnetic calorimetry)
energy resolution is assumed to be $3\%/\sqrt{E}\oplus 0.5\%$. In all
these, $E$ is the energy in~GeV units. We use the
cone-type Isajet \cite{isajet} jet-finding algorithm to group the
hadronic final states into jets. Jets and isolated leptons are defined as
follows:
\bi
\item Jets are hadronic clusters with $|\eta| < 3.0$,
$R\equiv\sqrt{\Delta\eta^2+\Delta\phi^2}\leq0.4$ and $E_T(jet)>40$~GeV.
\item Electrons and muons are considered isolated if they have $|\eta| < 2.5$, 
$p_T(l)>10 $~GeV with visible activity within a cone of $\Delta
R<0.2$ about the lepton direction, $\Sigma E_T^{cells} < \min[5,0.15p_{T}(l)]$~GeV.
\item We identify hadronic clusters as $b$-jets if they contain a $B$
hadron with $E_T(B)>$ 15~GeV, $|\eta(B)|<$ 3.0 and $\Delta R(B,jet)<$
0.5. We assume a tagging efficiency of 60$\%$ and light quark and gluon
jets can be mis-tagged as a $b$-jet with a probability $1/R_b$, with
$R_b=150$ for $E_{T} \leq$ 100~GeV, $R_b=50$ for $E_{T} \geq$ 250~GeV, and
a linear interpolation in between.
%for 100~GeV $\leq E_{T} \leq$ 250~GeV.  
\ei
We require the following cuts on our signal and background event
samples:
 %to extract those with a \break $\ell^\pm\ell^\pm +\eslt$ topology:
%                                                                                                        
\bi
\item {\it exactly} 2 isolated same-sign leptons with $p_T(\ell_1 )>20$~GeV
and $p_T(\ell_2 )>10$~GeV,
\item $n(b-jets)=0$ (to aid in vetoing $t\bar{t}$ background).
\ei
At this point the event rate is dominated by $WZ$ and $t\bar{t}$
backgrounds.  To reduce these further, we construct the transverse mass
of each lepton with $\eslt$ and require:
\bi
\item $\mtmin\equiv \min\left[ m_T(\ell_1,\eslt),m_T(\ell_2 ,\eslt )\right] > 125 \
{\rm GeV}$,
\ei
since the signal gives rise to a continuum distribution, while the
background has a kinematic cut-off around $\mtmin \simeq M_W$ (as long as the  
$\eslt$ dominantly arises from the leptonic decay of a single $W$).
After these
cuts, we are unable to generate any background events from $t\bar{t}$
and $WZ$ production, where the 1 event level in our simulation was
0.05~fb and 0.023~fb, respectively.  The dominant SM background for
large $\mtmin$ then comes from $Wt\bar{t}$ production for which we
find (including a QCD $k$-factor $k=1.18$ extracted from
Ref.~\cite{garzelli}) a cross section of $0.019$ ($0.006$)~fb after
the harder cuts, $\mtmin > 125$~(175)~GeV and $\eslt>200$~GeV that serve
to optimize the signal reach for high $m_{1/2}$ values.\footnote{We have
ignored detector-dependent backgrounds from jet-lepton misidentification
in our analysis, but are optimistic that these can be controlled by the
$\mtmin$ and $\eslt$ cuts.}

\begin{figure}[tbp]
\postscript{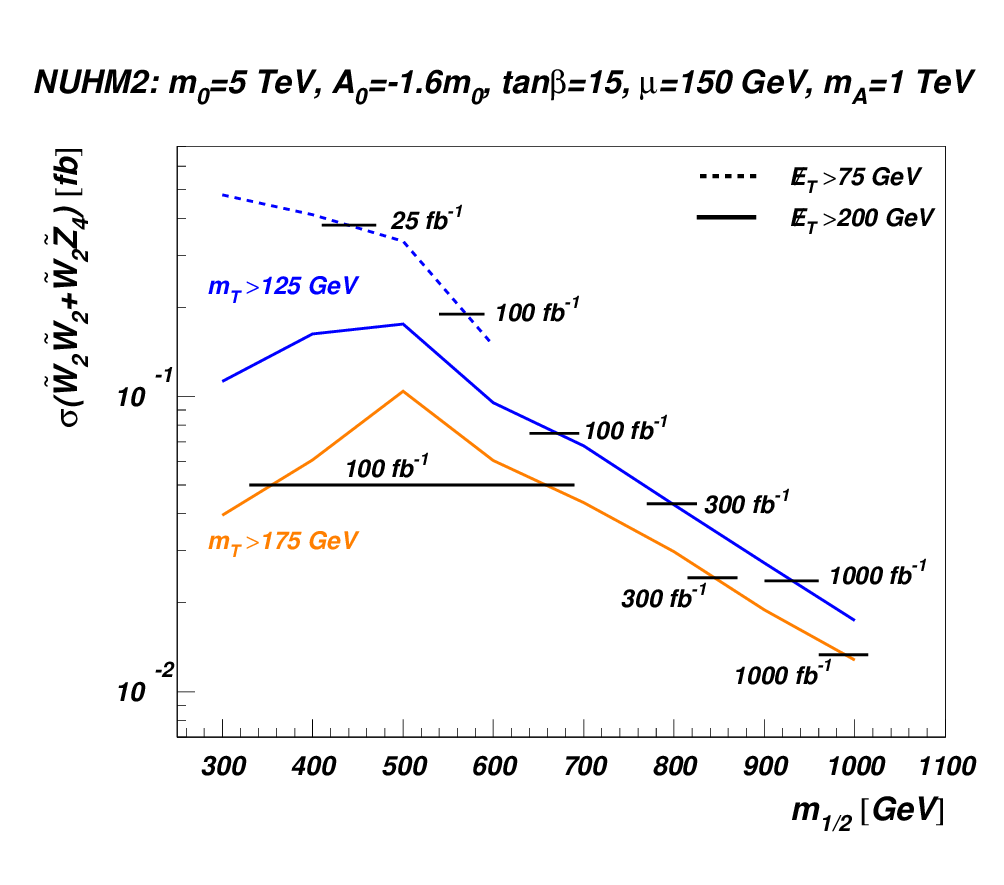}{1.1}
\caption{Same-sign dilepton cross section (fb) after cuts vs. $m_{1/2}$
along the RNS model line from just $\tw_2^\pm\tz_4$ and
$\tw_2^\pm\tw_2^\mp$ production with $t\bar{t}W$ background and
calculated reach for 100, 300 and 1000~fb$^{-1}$.  The upper solid and
dashed (blue) curves requires $\mtmin >125$~GeV while the lower solid
(orange) curve requires $\mtmin >175$~GeV. The signal is observable
above the horizontal lines.
\label{fig:ss}}
\end{figure}

The calculated signal rates after cuts along the RNS model line from
just $\tw_2^\pm\tz_4$ and $\tw_2^\pm\tw_2^\mp$ production are shown
vs. $m_{1/2}$ in Fig.~\ref{fig:ss} where the upper (blue) curves require
$\mtmin>125$~GeV and the lower (orange) curve requires $\mtmin>175$~GeV.  
The $\tw_2\tz_4$ and $\tw_2\tw_2$ cross sections are normalized to
those from Prospino\cite{prospino}.  For observability with an assumed
value of integrated luminosity, we require: 1)~significance $> 5\sigma$,
2)~Signal/BG$>0.2$ and 3)~at least 5 signal events.  The LHC reach for
SS diboson events for integrated luminosity values 100, 300 and 1000~fb$^{-1}$ 
is shown by horizontal lines in Fig.~\ref{fig:ss} and also in
Table~\ref{tab:reach}.  For just 10~fb$^{-1}$ of integrated luminosity
there is no LHC14 reach for SS dibosons,
while $\tg\tg$ production gives a reach of $m_{\tg}\sim 1.4$~TeV\cite{bblt1014}.  
However, for 100~fb$^{-1}$ the LHC14 reach for SS
dibosons extends to $m_{1/2}\sim 680$~GeV corresponding to $m_{\tg}\sim
1.6$~TeV in a model with gaugino mass unification.
The direct search for $\tg\tg$ gives a projected reach of $m_{\tg}\sim
1.6$~TeV~\cite{bblt2012}, so already the SS diboson signal offers a
comparable reach.  For 300~(1000)~fb$^{-1}$ of integrated luminosity, we
find the LHC14 reach for SS dibosons extends to $m_{1/2}\sim 840$~(1000)~GeV, 
corresponding to a reach in $m_{\tg}$ of 2.1 and 2.4~TeV.  These
numbers extend well beyond the LHC14 reach for direct gluino pair
production\cite{bblt1014}.
\begin{table}
\begin{center}
\begin{tabular}{|l|r|r|r|}
\hline
 Int. lum. (fb$^{-1}$) & $m_{1/2}$ (GeV) & $m_{\tg}$ (TeV) &  $m_{\tg}$ (TeV) [$\tg\tg$] \\
\hline
\hline
10   & -- & --    & 1.4 \\
100  & 680 & 1.6  & 1.6 \\
300  & 840 & 2.1  & 1.8 \\
1000 & 1000 & 2.4  & 2.0 \\
\hline
\end{tabular}
\caption{Reach of LHC14 for SUSY assuming various integrated luminosity values.
The reach is given for $m_{1/2}$ along the RNS model line, and also for the
equivalent reach in $m_{\tg}$ assuming heavy squarks. 
The corresponding reach in $m_{\tg}$ from $\tg\tg$ searches is also shown
for comparison.
\label{tab:reach}}
\end{center}
\end{table}

We emphasize here that the SS diboson signal from SUSY models with light
higgsinos is quite distinct from the usual SS dilepton signal arising
from gluino pair production, which is usually accompanied by numerous
hard jets and high $\eslt$. For instance, recent CMS searches for SS
dileptons from SUSY\cite{cms_ss} required the presence of two tagged
$b$-jets or large $H_T$ in the events; these cuts reduce or even eliminate
our SS diboson signal. Likewise, the cuts $n_j\ge 4$ high $p_T$ jets
along with $\eslt >150$~GeV required by a recent Atlas search for SS
dileptons from gluinos\cite{atlas_ss} would have eliminated much of the
SS diboson signal from SUSY with light higgsinos.

{\it Summary:} In SUSY models with light higgsinos, as motivated by 
electroweak naturalness considerations) 
the production of wino pairs gives rise to a novel
same-sign diboson plus modest hadronic activity signature. 
For an integrated luminosity of 100 (1000)~fb$^{-1}$ this SS
diboson signal should be observable at LHC14 for wino masses up to 550 (800)~GeV. 
Assuming gaugino mass unification, this extends the LHC SUSY reach
well beyond that of conventional searches for gluino pair production in
the case where squarks are heavy.

{\it Acknowledgements:} 
We thank Baris Altunkaynak for discovering an error in Fig. 2
in the early version of this paper.
We thank Andre Lessa for discussions.
This work was supported in part by the US Department of Energy, Office
of High Energy Physics, by Suranaree University of Technology, and by
the Higher Education Research Promotion and National Research University
Project of Thailand, Office of the Higher Education Commission.

%%%%%%%%%%%%%%%%%%%%%%%%%%%%%%%%%%%%%%%%%%%%%%%%%%%%%%

%
\end{document}